\documentclass[twocolumn,preprintnumbers,nofootinbib
]{revtex4}
\usepackage{amsmath,amssymb}

\newcommand{\eqa}{\begin{eqnarray}}
\newcommand{\neqa}{\end{eqnarray}}
\newcommand{\be}{\begin{equation}}
\newcommand{\ee}{\end{equation}}

\renewcommand{\texttt}{{}}
\usepackage{graphicx}
\usepackage{enumerate}
\newcommand{\beq}{\begin{equation}}
\newcommand{\eeq}{\end{equation}}
\newcommand{\beqa}{\begin{eqnarray}}
\newcommand{\eeqa}{\end{eqnarray}}
\newcommand{\nn}{\nonumber}
\begin{document}

\title{Entropic corrections to Newton's law} 
\author{Leonardo Modesto\footnote{email: lmodesto@perimeterinstitute.ca} \& Andrew Randono\footnote{email: arandono@perimeterinstitute.ca}}

\affiliation{Perimeter Institute for Theoretical Physics, 31 Caroline St., Waterloo, ON N2L 2Y5, Canada}

\date{\small\today}

\begin{abstract} \noindent
It has been known for some time that there is a deep connection between thermodynamics and gravity, with perhaps the most dramatic implication that the Einstein equations can be viewed as a thermodynamic equation of state. Recently Verlinde has proposed a model for gravity with a simple statistical mechanical interpretation that is applicable in the non-relatvistic regime. After critically analyzing the construction, we present a strong consistency check of the  model. Specifically, we consider two well-motivated corrections to the area-entropy relation, the log correction and the volume correction, and follow Verlinde's construction to derive corrections to Newton's law of gravitation. We show that the deviations from Newton's law stemming from the log correction have the same form as the lowest order quantum effects of perturbative quantum gravity, and the deviations stemming from the volume correction have the same form as some modified Newtonian gravity models designed to explain the anomalous galactic rotation curves.

\end{abstract}

\maketitle
\section*{Introduction: thermodynamic gravity}
The deep connection between thermodynamics and relativity first emerged over 35 years ago, with the pioneering work of Bekenstein, Hawking, and Unruh \cite{Bekenstein}\cite{Hawking}\cite{Unruh}. Since then the interrelation has been refined and greatly expanded. To date, the thermodynamic nature of causal horizons has, arguably, the strongest theoretical underpinnings of all studied quantum gravity effects. Viewed from a statistical mechanical perspective, the surmised existence of microscopic degrees of freedom that give rise to the thermodynamic laws in an appropriate limit has been the driving force of most quantum gravity research in the last few decades.

Beyond the initial work of Bekenstein and Hawking, perhaps the most surprising revelation relating thermodynamics and general relativity, is the discovery of Jacobson that the Einstein equations themselves can be viewed as nothing more than a thermodynamic equation of state under a set of minimal assumptions involving, most notably, the equivalence principle and the identification of the the area of a causal horizon with its entropy \cite{Jacob}\cite{wald}\cite{Pad}\cite{CM}. Recently Verlinde \cite{ver} has put forward an idea similar in spirit to Jacobson's thermodynamic derivation of the Einstein equations, where it is argued that Newton's law of gravitation can be understood as an entropic force in direct analogy with the well-known (and experimentally verified) entropic explanation of the thermodynamically emergent forces on polymers immersed in a heat bath. As a relatively new idea, Verlinde's entropic argument for the origin of Newtonian gravity has its strengths and weaknesses, which we will discuss more thoroughly in section \ref{Weaknesses}. Whereas Jacobson's derivation is logically clear and theoretically sound, the statistical mechanical origin of the thermodynamic nature of general relativity remains obscure. In contrast, Verlinde's explanation of Newton's law of gravitation at the very least offers a strong analogy with a well understood statistical mechanism. In addition, the derivation opens the door for deviations to ordinary gravity in a weak field, non-relativistic limit, which is the most accessible observational regime for testing quantum gravity phenomenology \cite{PhQG}.

With this in mind, we will follow the logic of Verlinde's derivation while considering deviations to the area law. More specifically the deviations to the entropy we will consider are of the form 
\begin{eqnarray}
S(A) = \frac{A}{4 l_P^2} - a \log \left(\frac{A}{l_P^2} \right) + b \left( \frac{A}{l_P^2} \right)^{3/2}
\label{Gentropy} 
\end{eqnarray}
for constants $a$ and $b$ on the order of unity or less. Our primary motivation for considering these deviations from the standard entropy is retrospective in light of the results of section \ref{corre}, where we show that following Verlinde's procedure these deviations produce compelling corrections to Newton's laws. On the other hand, the extra terms are well-motivated by previously proposed corrections to the area law stemming from bottom-up, non-perturbative frameworks of quantum gravity. The log correction appears to have a {\it universality}, having been emerged from the counting of microstates in many independent quantum gravity theories (see e.g. \cite{CarlipUniversailty}\cite{Major}). In this paper we show that the entropic origin of Newton's laws offer an explanation for this universality -- the resultant corrections to Newton's laws stemming from the log correction to the entropy following Verlinde's entropic argument are identical up to the choice of constant $a$ to the lowest order non-relativistic corrections to Newton's laws that follow from perturbative quantum gravity. The remaining term, the ``volume" correction to the area law, is also motivated by a model for the microscopic degrees comprising the black hole entropy in Loop Quantum Gravity \cite{BHent}, where an $A^{3/2}$ correction to the area law emerged. Again following Verlinde's entropic argument, we will show (up to the choice of constant $b$) this term gives rise to the $1/R$ correction to Newton's laws that is commonly invoked in modified Newtonian gravity theories as an explanation for the observed anomalous galactic rotation curves.

We offer the following perspective for interpretation of these results. With its relatively loose and pedagogical style, it is easy to poke holes in Verlinde's argument. Indeed, the objection has been raised that the main content of the derivation may simply be the by-product of consistent dimensional analysis. We offer here that the association of the log correction to the entropy law and the perturbative graviton mediated correction to Newton's law we derive can be viewed as a strong consistency check on validity of the idea, while at the same time offering a compelling explanation for the apparent universality of the log correction. In addition, the association of the volume correction to the entropy with a well-known model for the anomalous galactic rotation curves suggests that the entropic explanation of Newtonian gravity could be a powerful tool for constructing verifiable or falsifiable quantum gravity inspired corrections to ordinary gravity in a regime where these corrections are observationally accessible. 

With this in mind, we begin the section \ref{Weaknesses} with a detailed outline of the strengths and weaknesses of Verlinde's argument with the goal of isolating the points where the argument should be improved. Following this we derive corrections to Newton's laws in section \ref{corre}.

\section{Newtonian gravity as an entropic force} \label{Weaknesses}
The picture Verlinde has put forward is an attempt to model non-relativistic gravity as a force originating from an entropy gradient, analogous to the known entropic forces on polymers immersed in a heat bath
\cite{CINA0}. In the latter case, the force that originates purely from the tendency of a closed system to evolve in the direction increasing entropy obeys the simple {\it thermodynamic} equation $F\sim T\frac{\Delta S}{\Delta x}$. Modelling the entropy gradient for polymers in simple cases leads to Hook's law. One key property of the force is the temperature dependence, which can be viewed as one of the defining feature of entropic forces. Verlinde's program extends this idea to the gravitational force between test masses using a set of assumptions loosely motivated by concepts that have emerged from the study of the relation between thermodynamics and relativity. Here we will summarize systematically what we perceive to be the key assumptions in Verlinde's entropic derivation of Newton's laws.

Suppose we have two masses one a test mass and the other considered as the source with respective masses $m$ and $M$. From Newton's third law, it should be expected that the roles of the test mass and source can be swapped without changing the content of the argument, but we will stick with these labels. Centered around the source mass $M$, is a spherically symmetric surface $\mathcal{S}$ which will be defined with certain properties that will be made explicit later. To derive the entropic law, the surface $\mathcal{S}$ is between the test mass and the source mass, but the test mass is assumed to be very close to the surface as compared to its (reduced) Compton wavelength $\lambda_m=\frac{\hbar}{mc}$. Verlinde then imposes the following assumptions (taking some liberty in presentation)

\begin{enumerate}
\renewcommand{\theenumi}{\roman{enumi}}
\renewcommand{\labelenumi}{\theenumi}

\item\label{Entropy}{Close to the surface $\mathcal{S}$, as compared to the Compton wavelength $\lambda_m$, the change in entropy of the surface is proportional to the change in radial distance of the test mass from the surface. The constant of proportionality is fixed by
\beq
\Delta S=2\pi k_B \frac{\Delta x}{\lambda_m}+\mathcal{O}(\left(\frac{\Delta x}{\lambda_m}\right)^2)
\eeq
where $k_B$ is Boltzmann's constant. This constant of proportionality is chosen such that the entropic enforce that ensues satisfies $F=ma$ when the temperature is identified with the Unruh temperature of the causal horizon of a congruence of accelerated observers.}

\item\label{SurfaceEnergy}{The energy of the surface $\mathcal{S}$ is identified with the relativistic rest mass of the source mass:
\beq
E_{\mathcal{S}}=Mc^2.
\eeq
}
\item\label{SurfaceDOF}{On the surface $\mathcal{S}$, there live a set of ``bytes" of information that scale proportional to the area of the surface so that
\beq
N=\frac{A_{\mathcal{S}}}{\ell_p^2}
\eeq
where $\ell_p=\sqrt{\frac{G\hbar}{c^3}}$ is the Planck length.}

\item\label{Equipartition}{The surface is in thermal equilibrium at a temperature $T$, and the energy of the surface $\mathcal{S}$ is equipartitioned among the bytes on the surface so that
\beq
E_{\mathcal{S}}=N\cdot \frac{1}{2}k_B T\,.
\eeq
}

\item\label{EOS}{Finally it is assumed that the force on the particle follows from the generic form of the entropic force governed by the thermodynamic equation of state
\beq
F\Delta x=T\Delta S\,.
\eeq}

\end{enumerate}
With these assumptions, it is a simple matter of algebraic manipulation to derive Newton's law of universal gravity, $F=ma=GMm/r^2$.

Let us now examine these assumptions critically. We first point out an apparent tension between (\ref{Entropy}) and (\ref{SurfaceDOF}). Both assumptions define an entropy change associated with the surface that appear to be in conflict. Although it is not specified directly in the paper, (\ref{SurfaceDOF}) makes the implicit assumption that the entropy of the surface is proportional to the area. For example, suppose each ``byte" of information comprising the surface degrees of freedom is itself comprised of $n$ ``bits" as is common in many models of horizon entropy. Then the total entropy of the surface would be $S_\mathcal{S}=\log(n^N)=N\, \log(n)\sim A_{\mathcal{S}}$. This is further supported by the equipartition assumption (\ref{Equipartition}) from the general principle that it is the microscopic degrees of freedom that comprise the entropy of the surface that each obtain $\frac{1}{2}k_BT$ of energy on average in thermal equilibrium. Thus, there are two seemingly competing definitions of entropy in the construction one the scales like the distance from the horizon $\Delta S \propto \Delta x$ and one that scales like the area $\Delta S \propto \Delta A$. These different scalings are not necessarily mutually incompatible, but it is not obvious how they {\it are} compatible. The authors are currently investigating a polymer inspired model whereby the two scalings are transparently compatible, but it is not yet clear how this applies to the case of gravity. In this paper, we will skirt this issue following the logic of a follow-up work by Smolin \cite{LS}. There, he considered the information theoretic change in entropy has the flux of a bit or byte across the surface $\mathcal{S}$, which is necessarily discretized. The entropy is assumed to change by one fundamental unit when the mass $m$ is at a distance from the surface given by $\Delta x=\beta\lambda_m$, where $\beta$ is a dimensionless constant on the order of one that will be fixed later to give the proper coupling constant of gravity. The logic is in keeping with Bekenstein's argument that when the particle distance is on the order of one Compton wavelength, it is indistinguishable from the horizon itself (this argument carries more weight when $\mathcal{S}$ is a horizon comprised of a congruence of null generators, recalling that the (non-reduced) Compton length is the wavelength a photon would have were its energy equal to the rest mass of the particle).

A potentially more serious objection to the construction revolves around the peculiar combination of non-relativistic and relativistic concepts. More specifically, the definition of the surface $\mathcal{S}$ appear to be, at present, ill-defined. From assumption (\ref{SurfaceEnergy}) the energy of the surface is the energy associated with the rest mass of the source mass $M$, suggesting that the surface is fundamentally connected with the source mass. This in itself is potentially a problem since, in the non-relativistic scenario, typically the thermodynamic free-energy only consists of energies above the rest mass energy. Barring this rather weak objection, it is natural to think of the surface $\mathcal{S}$ as a horizon associated with the mass $M$, and many of the properties ascribed to this surface are properties that one would only expect from a horizon, including assigning the rest mass energy $Mc^2$ to the surface. However, since the construction is intrinsically non-relativistic there is no obvious horizon present in the physical scenario. One may argue that even in the non-relativistic limit there is a remedial concept of a horizon if one impose a velocity limit $v_{max}$. In this case, the classical unbound states must satisfy $\frac{1}{2}mv^2 \geq \frac{GmM}{r}$ which gives a minimal radius for a free state given by $r_0=\frac{2GM}{v_{max}^2}$. Setting the maximum velocity numerically equal to $c$ yields precisely the Schwarzschild radius. On the other hand, in this case, we still face the problem that it is precisely the Schwarzschild radius where relativistic effects dominate and Newton's laws are no longer a good approximation. One may argue that one can extend the dependence of entropy (\ref{Entropy}) to regions well outside the surface $\mathcal{S}$. Setting aside that this is a very strong restriction placed on the entropy dependence, one is still faced with the problem that in the derivation one must set the area of the surface equal to $4\pi R^2$, where $R$ is initially the radius of the surface in this equality, but is later identified with the radial distance of the particle from the point mass in the derived Newton's law. Thus, the derivation itself only makes sense when the particle is close to the surface $\mathcal{S}$ so that $R_{\mathcal{S}}\approx R_{m}$. Furthermore, the existence of a the Schwarschild horizon is dependent on density, not mass, and Newton's law holds equally well outside a low density distribution of matter where no horizon is present. 

Faced with these difficulties, one may try to abandon the notion of a horizon and take $\mathcal{S}$ to be an ordinary surface, the energy of which is identified with the total energy contained in in the interior. The surface degrees of freedom are unspecified bits of information, the content of which is preserved as the area of the surface expands or contracts. However, in this case, it would be difficult to justify equations (\ref{SurfaceDOF}) and (\ref{Equipartition}) simultaneously. Since the energy of the surface is associated with the total energy contained in the volume, the equipartition theorem states that the free energy is distributed equally among the degrees of freedom associated with the surface. However, if $\mathcal{S}$ is not a horizon, or more generally a surface that hides information inaccessible to observers in the exterior, one expects that the number of degrees of freedom associated with the surface should scale like volume bounded by the surface, not the area -- in this case, the equipartition theorem would assign $\frac{1}{2}k_BT$ to degrees of freedom whose number scales like the volume not the area. The salient point here is that it is a very special feature of surfaces that hide information from the outside world that their degrees of freedom scale like the area (as do the degrees of freedom stemming from the entropy of a causal horizon, or the entanglement entropy associated with a region), not like the volume (as do the degrees of freedom of quantum field theory or standing waves confined to a box).  

Faced with these difficulties one may be tempted to abandon the entropic picture of gravity as a mere coincidence, perhaps stemming from consistent dimensional analysis. Rather than attack these problems directly, we will offer a strong consistency check of the construction, and a simple model whereby plausible deviations to Newton's law can be obtained by tweaking Verlinde's procedure. This should serve to strengthen the viability of the model of gravity as an entropic force, but should also be viewed as a call to the community to strengthen the arguments employed in the construction.

\section{Generic entropic corrections to Newton's universal law of gravitation} \label{corre}
As mentioned in the introduction, the strength of Verlinde's entropic explanation of Newtonian gravity is that it relies on an analogy with simple, intuitive thermodynamic processes in polymer physics. This gives a more clear statistical mechanical picture of the gravitational interaction while opening the door for deviations from Newton's law of universal gravitation in a regime that may be accessible by experiment. With this in mind, we consider a generic modification of the relation between the area of $\mathcal{S}$ and its entropy \cite{LQBHs} given by
\beq
S_{\mathcal{S}} = \frac{A c^3 k_B}{4 \hbar G} +  k_B s(A)\,.
\eeq
We will largely follow the spirit of the procedure outlined by Verlinde. The most obvious modification is the replacement of the linear relationship $\Delta S\sim \Delta x$ with the previously mentioned discretized version proposed by Smolin. We will first  consider the most general situation and later specialize to a specific modification $s(A)$ and discuss how the picture naturally fits in the context of Loop Quantum Gravity. 

Our modified list of assumptions:
\begin{enumerate}
\renewcommand{\theenumi}{\Roman{enumi}}
\renewcommand{\labelenumi}{\theenumi}

\item\label{Entropy 2}{When a test mass $m$ is a distance $\Delta x_0=\beta \lambda_m$ away from the surface $\mathcal{S}$, the entropy of the surface changes by one fundamental unit $\Delta S_0$ fixed by the discrete spectrum of the area of the surface via the relation
\beq
\Delta S = \Delta A \left( \frac{c^3}{4 \hbar G} + \frac{\partial s}{\partial A}\right) k_B.
\eeq  }

\item\label{SurfaceEnergy 2}{The energy of the surface $\mathcal{S}$ is identified with the relativistic rest mass of the source mass:
\beq
E_{\mathcal{S}}=Mc^2.
\eeq
}
\item\label{SurfaceDOF 2}{On the surface $\mathcal{S}$, there live a set of ``bytes" of information that scale proportional to the area of the surface so that
\beq
A =Q N
\eeq
where $N$ is an integer labeling the number of bytes and $Q$ is the fundamental ``charge" or area gap (since $\Delta A\big|_{N=1}=Q$) determined by the microscopic theory.}

\item\label{Equipartition 2}{The surface is in thermal equilibrium at a temperature $T$, and the energy of the surface $\mathcal{S}$ is equipartitioned among the bytes on the surface so that
\beq
E_{\mathcal{S}}=\alpha N\cdot \frac{1}{2}k_B T\,.
\eeq
The constant $\alpha$ is included here to account for different possible values of the average number of bits per byte.}

\item\label{EOS 2}{Finally it is assumed that the force on the particle follows from the generic form of the entropic force governed by the thermodynamic equation of state
\beq
F=T\frac{\Delta S}{\Delta x}
\eeq}
where it is understood that $\Delta S$ is one fundamental unit of entropy when $|\Delta x|=\beta \lambda_m$, and the entropy gradient points radially from the outside of the surface to inside.
\end{enumerate}

Although we have closely followed Verlinde's theory and have not adopted a specific model for the microscopic theory, these assumptions naturally fit within the framework of Loop Quantum Gravity with some reasonable assumptions. First we identify the surface $\mathcal{S}$ as a closed ball whose macroscopically available degrees of freedom that are available to an external observer are completely characterized by the volume, $V_0$ and the surface area, $A_0$, of the region. The microscopic degrees of freedom comprising the entropy, understood as a measure of ignorance of the external observer, are the spin network configurations consistent with this set of macroscopic data. The area of the surface is well known as a quantum operator, and depends only on the punctures on the surface where a labelled spin network edge enters the region. Thus, some of the microscopic degrees of freedom are distinguished by the different ways of puncturing the surface with labelled edges such that the total area is $A_0$. For large $A_0$, a typical configuration is dominated by  spin-$\frac{1}{2}$ representations, so for simplifications we will consider configurations where all the punctures are spin-$\frac{1}{2}$. Following \cite{BHent}, the remaining degrees of freedom can be characterized by the space of intertwiners of the surface, defined as irreducible $SU(2)$-invariant representation of the tensor product of the representations of the labelled punctures. The total entropy of the surface $\mathcal{S}$ can then be computed by calculating the dimension of the space of intertwiners. The key result relevant for this work is that when this procedure is carried out, it is found that the entropy to lowest order scales linearly with the area, but there are higher order corrections of the form $s[A]$, which we will discuss in more detail in the next section. Under the assumption that all spins are spin-$\frac{1}{2}$ representations, $N$ is identified with the number of punctures, and the parameter $Q$ in (\ref{SurfaceDOF 2}) is given by the area gap $\Delta A_0=8\pi \ell_P^2 \gamma  \sqrt{\frac{1}{2}(\frac{1}{2} +1)}$, where $\gamma$ is the Immirzi parameter fixed in LQG by the requirement that the entropy of an isolated horizon agrees at lowest order with the Bekenstein-Hawking entropy. In addition, if all punctures are spin-$\frac{1}{2}$, there are $2$ bits per byte so $\alpha=2$. Returning to assumption (\ref{Entropy 2}), in LQG one can model fermionic degrees of freedom by a free edge with one end connected to a vertex of the spin network (see Fig. 1). Taking this to be the rest mass, the distance of the spinor test mass from the surface can be identified as the expectation value of a complicated spin-network operator defining the distance. When expectation value $\langle \Delta x \rangle=\beta \lambda_m$, the surface is deformed to capture the free edge as a new puncture. Since this puncture is itself an edge labelled by as spin-$\frac{1}{2}$ representation, the new puncture adds a quanta of area equal to the minimal area gap $\Delta A_0$, which in turn fixes $\Delta S_0=\Delta A_0\left( \frac{c^3}{4 \hbar G} + \frac{\partial s}{\partial A}\right) k_B$. Thus, the assumptions (\ref{Entropy 2})-(\ref{EOS 2}) naturally fit within the framework of LQG.

\begin{figure}
 \begin{center}
\includegraphics[height=4.0cm]{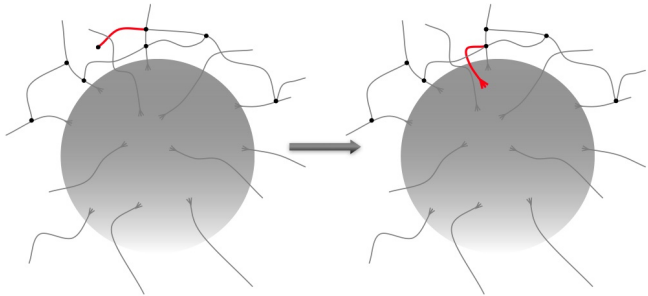}
   \end{center}
  \caption{\label{BHSN} In this picture the increase of the area upon inclusion of matter to the surface is explained adding a fermion to the 
  graph, represented  by a red link in the spin-network. Dynamical evolution moves the 
  fermion degree of freedom closer to the surface, represented by a vertex jump. The hypothesis is that its distance from $\mathcal{S}$ is $\langle \Delta x \rangle \approx \lambda_m$ by quantum uncertainty the particle is indistinguishable from the surface. Thus, in this picture the surface must be deformed to include the edge as a new puncture, as shown on the right.}
  \end{figure}

Setting LQG aside for the moment, it is a simple matter in the generic setting to determine the modification to Newton's law of gravitation that the corrections, $s(A)$, to the entropy add. We first solve (\ref{SurfaceEnergy 2}), (\ref{SurfaceDOF 2}), and (\ref{Equipartition 2}) simultaneously for the temperature, identifying $A=4\pi R^2$ where $R$ is approximately the radius of the test mass from the source to yield:
\beq
T=\frac{M}{R^2}\frac{Q c^2}{2\pi k_B\alpha}\,.
\eeq
The fundamental unit of entropy is fixed by $\Delta S_0 = Q \left( \frac{c^3}{4 \hbar G} + \frac{\partial s}{\partial A}\right) k_B$. Thus in total, the equation of state (\ref{EOS 2}) yields
\beq
F=-\frac{Mm}{R^2} \left(\frac{Q^2 c^3}{2\pi  \,\hbar \,\alpha\beta}\right)\left[\frac{c^3}{4\hbar G}+\frac{\partial s}{\partial A} \right]_{A=4\pi R^2}
\eeq
which yields Newton's laws to first order if and only if $Q^2=8\pi \, \alpha\beta \,\ell_P^4$. With this identification, the modified Newtonian force is
\beq
F=-\frac{GMm}{R^2}\left[1+4\ell_P^2 \frac{\partial s}{\partial A} \right]_{A=4\pi R^2}
\eeq

\subsection{Specific quantum corrections to Newton's law of gravitation}
Let us now specialize to a specific form for the corrected area-entropy relation given by
\beq
S[A]=\frac{A k_B}{4 \ell_P^2}-a k_B \ln{\left(\frac{A}{\ell_P^2}\right)}+b k_B  \left(\frac{A}{\ell_P^2}\right)^{\frac{3}{2}}\,. \label{entropy2}
\eeq
As previously mentioned, these corrections are well motivated from bottom-up quantum gravity theories. The log correction to the area-entropy relation appears to have an almost universal status, having been derived from multiple different approaches to the calculation of entropy from counting microscopic states in different quantum gravity models \cite{CarlipReview}. The ``volume" correction is physically natural in the sense that the ordinary degrees of freedom (bytes) of quantum fields generically scale like the volume as opposed to the area. Thus, this correction can be interpreted as a interpolation between a holographic phase of gravity, and a field theoretic phase. Furthermore, specific models of black hole entropy in LQG derive precisely a correction of this form \cite{ET} (see section \ref{MOND}). 

The modified force law and potential ($V=-\int F dR$) obtained from the procedure are
\beqa
F&=& -\frac{GMm}{R^2}\left[1-a\frac{\ell_P^2}{\pi R^2} +b\,12\sqrt{\pi}\,\frac{R}{\ell_P}\right],  \label{VLV}\\
V&=&-GMm\left[\frac{1}{R}-a\frac{\ell_P^2}{3\pi R^3}-b\,\frac{12\sqrt{\pi}}{\ell_P}\,\log{\left(\frac{R}{\ell}\right)}\right], \nn 
\eeqa
where in the last line we have absorbed the constant of integration into an unspecified length parameter $\ell$. Let us now analyze the two corrections separately.

\subsection{Log correction: perturbative quantum gravity}
First we consider the corrections to Newton's law stemming from the log correction to the area-entropy relation. The key observation is that the correction so obtained has precisely the same functional form as the correction one obtains from perturbative quantum gravity. 

As mentioned previously, the log correction has a near universal status, having been derived by many different quantum gravity models. Aside from the many bottom-up models, the correction also emerges as a perturbative quantum correction as demonstrated by Fursaev \cite{FU}, where the entropy of a Schwarzschild black hole is calculated considering the corrections coming from one-loop effects of quantum matter fields near the black hole. 
Defining $N_s$ the number of massless fields of spin $s$ present in the quantum field theory
they obtain
\begin{eqnarray}
&& S = \frac{A}{4} + \sigma \log A, \label{LOGSQFT}  \\
&& \sigma =  \frac{1}{90}\left(N_0 + \frac{7}{4} N_{1/2} - 13 N_1 - \frac{233}{4} N_{3/2} + 212 N_2\right). 
\nonumber 
\end{eqnarray}
Given the $\log$-correction to the entropy we obtain a quantum correction to the gravitational 
potential which is compatible with the one-loop correction to the Newton law in perturbative
quantum gravity.

One can also calculate the one-loop corrections to the gravitational force law in the perturbative framework. The correction to the non-relativistic potential from an approximation of the full amplitude proceeds as follow. First, the the relation between the expectation value of the $S$-matrix to the Fourier
transform of the potential $\tilde{V}(q)$ is
\begin{eqnarray}
\langle k_1, k_3| S | k_2, k_4 \rangle = - 2 \pi i \tilde{V}(\vec{q}) \delta(E_i - E_f),
\label{SPQG}
\end{eqnarray}
where $k_1$, $k_3$ and $k_2$, $k_4$ are the ingoing and outgoing momentum respectively,
$q = k_2 - k_1= k_3 - k_4$ and $E_i- E_f$ is the energy difference between the incoming and 
outgoing states. From the diagrammatic expansion we find 
\begin{eqnarray}
\langle k_1, k_3| S | k_2, k_4 \rangle = (2 \pi)^2 \delta(k_2+k_4-k_1-k_3) (i {\mathcal M}),
\label{SPQG2}
\end{eqnarray}
where the general form of scattering amplitude ${\mathcal M}$ is 
\begin{eqnarray}
&& {\mathcal M} \approx (A + B q^2 + \dots + \frac{C_0 \kappa^4}{q^2} \nonumber \\
&& + C_1 \kappa^4 \log(- q^2) 
+\frac{C_2 \kappa^4 m}{\sqrt{- q^2}} + \dots ),
\label{M}
\end{eqnarray}
but only the $C_1$ and $C_2$ terms yield the leading quantum correction to the 
gravitational potential. In the non-relativistic limit ($q=(0,\vec{q})$) 
$\tilde{V}(\vec{q}) = {\mathcal M}/4 m_1 m_2$ and 
\begin{eqnarray}
V(r) = - \frac{1}{ 4 m_1 m_2} \int \frac{d^3 q}{(2 \pi)^3} e^{i \vec{q}\cdot \vec{r}} {\mathcal M}\,.
\label{VX}
\end{eqnarray}
The relevant term in our case comes from $C_1$ whose Fourier transform is given by the identity
\beq
\int \frac{d^3q}{(2\pi)^3} e^{i\vec{q}\cdot \vec{r}} \log(|\vec{q}|^2) =-\frac{1}{\sqrt{2\pi} |\vec{r}|^3}\,.
\eeq
Thus ,the first order non-relativistic quantum correction to the Newtonian potential is of the form 
\beq
V(R)=-GMm\left(\frac{1}{R}+...-\frac{a' \ell_P^2}{R^3}+...\right)
\eeq
where $a'$ is independent of $G$ and the masses, but does depend on the particle species included in the one-loop corrections. We can conclude that following Verlinde's procedure, the log-correction to the entropy yields perturbations to Newton's law with a functional form that are consistent with the deviations coming from perturbative quantum gravity. 

\begin{figure}
 \begin{center}
\includegraphics[height=5.5cm]{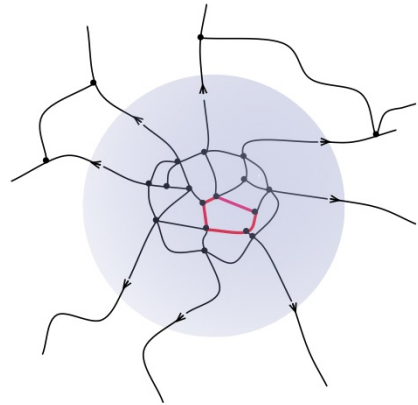}
   \end{center}
  \caption{\label{BHSN0} 
  The picture represents the spin-network graph when we have more loops 
  inside the region of surface $\mathcal{S}$. The red one is an example of loop.}
  \end{figure}

\subsection{Volume correction}
\label{MOND}
Let us now focus on the corrections to Newton's laws stemming from the volume correction to the area-entropy relation. We notice here that the form of the potential is of the same functional form as the correction to the Newtonian potential posited by Modified Newtonian Dynamics (MOND) in order to model the anomalously flat galactic rotation curves.
In the paper \cite{ET} Livine and Terno calculated the entropy in LQG
introducing a dependence on the number of loops $L$ for the 
spin-network state dual to the region of surface $\mathcal{S}$ (see Fig. 2). 
In the limit of a larger number of loops $L \gg N$ the entropy reads
$S_{L \gg N} \sim N \log L \sim N^{3/2} \propto A^{3/2}$,
where in the second equality we used $L \sim 2^{\sqrt{N}}$. 
%
The result obtained in \cite{ET} suggests to study the entropy correction 
$s(A) \sim  b k_B \left( A c^3/G \hbar \right)^{3/2}$ introduced in the previous section.
At large distances, the leading term in the force law is the modification due to the volume correction of the are-entropy relation. For a circular orbit, the acceleration is $\frac{v^2}{R}$, and the modified Newton's law gives
\beq
F=\frac{mv^2}{R}\approx \frac{GMm}{R^2}+b\frac{12\sqrt{\pi}}{\ell_P} \frac{GMm}{R}\,.
\eeq
For very large distances, the tangential velocity of a circular orbit approaches a constant 
\beq
v^2 \approx b \frac{12\sqrt{\pi} GM}{\ell_P}\,.
\eeq
Comparing this with the model employed by most modified Newtonian gravity theories  \cite{MOND} where the velocity approaches $v^2 = \sqrt{ G M a_o}$, we find that the results are compatible if 
\beq
b = \frac{\ell_P}{3}\sqrt{\frac{a_o}{16\pi G M}}\,.
\eeq
It remains to be seen if more extensive study of the volume correction to the are-entropy relation from microstate counting of a bottom-up theory can produce this scaling.
\section{Summary and Conclusions}
After critically analyzing the logical construction of Newtonian gravity as an entropic force, we have presented strong evidence in favor of the consistency of the model. In particular we have shown that an apparently universal log correction to the area-entropy, yields deviations from Newton's laws that are identical in form to those obtained from perturbative quantum gravity. This at once sheds light on the reason for the universality of the log correction and provides a strong consistency check on Verlinde's model. Somewhat more speculatively, we also considered the volume correction to the area-entropy relation, and found that the deviation from the Newton's law so obtained was similar in form to the proposed modification of Modified Newtonian Dynamics to explain the anomalous galactic rotation curves. Although it has yet to be seen whether these corrections to the entropy are robust predictions of a bottom-up quantum gravity model, and coupling has the appropriate mass scaling, the association does support lend support to the entropic picture of gravity as a fruitful model for obtaining verifiable or falsifiable corrections to known physical laws in a observationally accesible regimes.

\section*{Acknowledgements} 
Research at Perimeter Institute is supported by the Government of Canada
through Industry Canada and by the Province of Ontario
through the Ministry of Research \& Innovation. AR is supported by a grant from the NSF International Research Fellowship program, OISE0853116.

\end{document}